\documentclass[RNAAS]{aastex62}

\begin{document}

\title{The ultraviolet-infrared color-magnitude relation of star-forming galaxies}

\correspondingauthor{Michael J. I. Brown}
\email{Michael.Brown@monash.edu}

\author[0000-0002-1207-9137]{M.~J.~I. Brown}
\affil{School of Physics and Astronomy, Monash University, Clayton, Victoria 3800, Australia}
\author[0000-0002-2733-4559]{J. Moustakas}
\affil{Department of Physics and Astronomy, Siena College, 515 Loudon Road, Loudonville, NY 12211}
\author[0000-0002-4939-734X]{T.~H. Jarrett}
\affil{Astrophysics, Cosmology and Gravity Centre (ACGC), Astronomy Department, University of Cape Town, Private Bag X3, Rondebosch 7701, South Africa}
\author[0000-0002-9871-6490]{M. Cluver}
\affil{Centre for Astrophysics and Supercomputing, Swinburne
University of Technology, Hawthorn, VIC 3122, Australia}
\affil{Department of Physics and Astronomy, University of the
Western Cape, Robert Sobukwe Road, Bellville, 7535, South
Africa}

\section{} 

Galaxy bimodality, described by the red sequence and blue cloud, has been central to our understanding of galaxy evolution since the turn of this century \citep[e.g.,][]{2004ApJ...600..681B,2004ApJ...608..752B,2009ARA&A..47..159B}. Passive galaxies follow a narrow color-magnitude relation while star-forming galaxies in the blue cloud have a broader range of optical colors, resulting from a range of stellar populations, star formation rates and dust obscuration. Although star-forming galaxies are diverse, they do fall along well-established correlations with mass, including the mass-metallicity relation \citep[e.g.,][]{2004ApJ...613..898T}, star formation rate versus mass \citep[e.g.,][]{2007ApJ...660L..43N} and declining dust content with decreasing mass \citep[e.g.,][]{2006ApJ...639..157W}. These relations manifest themselves in broadband photometry, albeit outside the optical wavelength range, as illustrated by the dependence of infrared colors on galaxy type \citep[e.g.,][]{2011ApJ...735..112J}.

The far-ultraviolet and mid-infrared are both star formation rate tracers, with the former tracing massive stars while the latter traces blackbody emission from warm dust. While far-ultraviolet luminosity is directly proportional to star formation rate, for $\lesssim L^*$ galaxies mid-infrared luminosity is proportional to star formation rate to the power of $\sim 1.3$ \citep[e.g.,][]{2015AA...584A..87C,2017ApJ...847..136B}, which is a consequence of dust content varying with galaxy mass. We thus expect a blue sequence to be present in ultraviolet-infrared color-magnitude diagrams.

To measure the ultraviolet-infrared color-magnitude relation, we use the local galaxy sample of \citet{2014ApJS..212...18B, 2017ApJ...847..136B} and their multiwavelength matched aperture photometry (in AB magnitudes). We limit the sample to galaxies with $m_{W2}-m_{W3}>-0.5$, which excludes passive galaxies from the \citet{2014ApJS..212...18B} sample, and we remove active galactic nuclei with the emission line ratio criterion of \citet{2003MNRAS.346.1055K}. To correct the GALEX $FUV$ photometry for internal dust obscuration we use $A_{FUV}  \propto (M_{FUV} - M_{NUV})$, leaving the constant as a free parameter that we use to minimize the scatter of the color-magnitude relation. 

\begin{figure}
\begin{center}
\includegraphics[width=0.49\textwidth,angle=0]{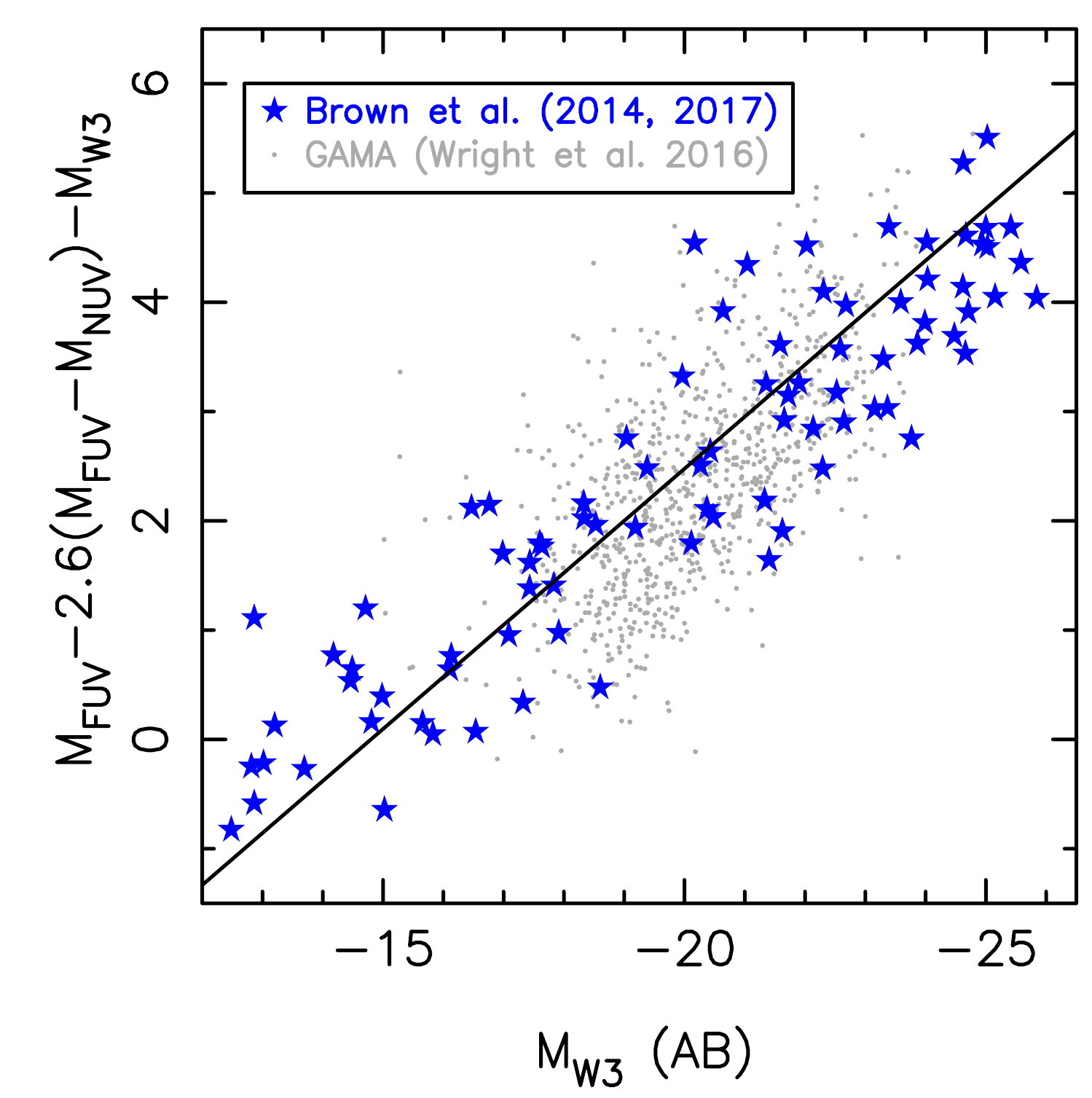}
\includegraphics[width=0.49\textwidth,angle=0]{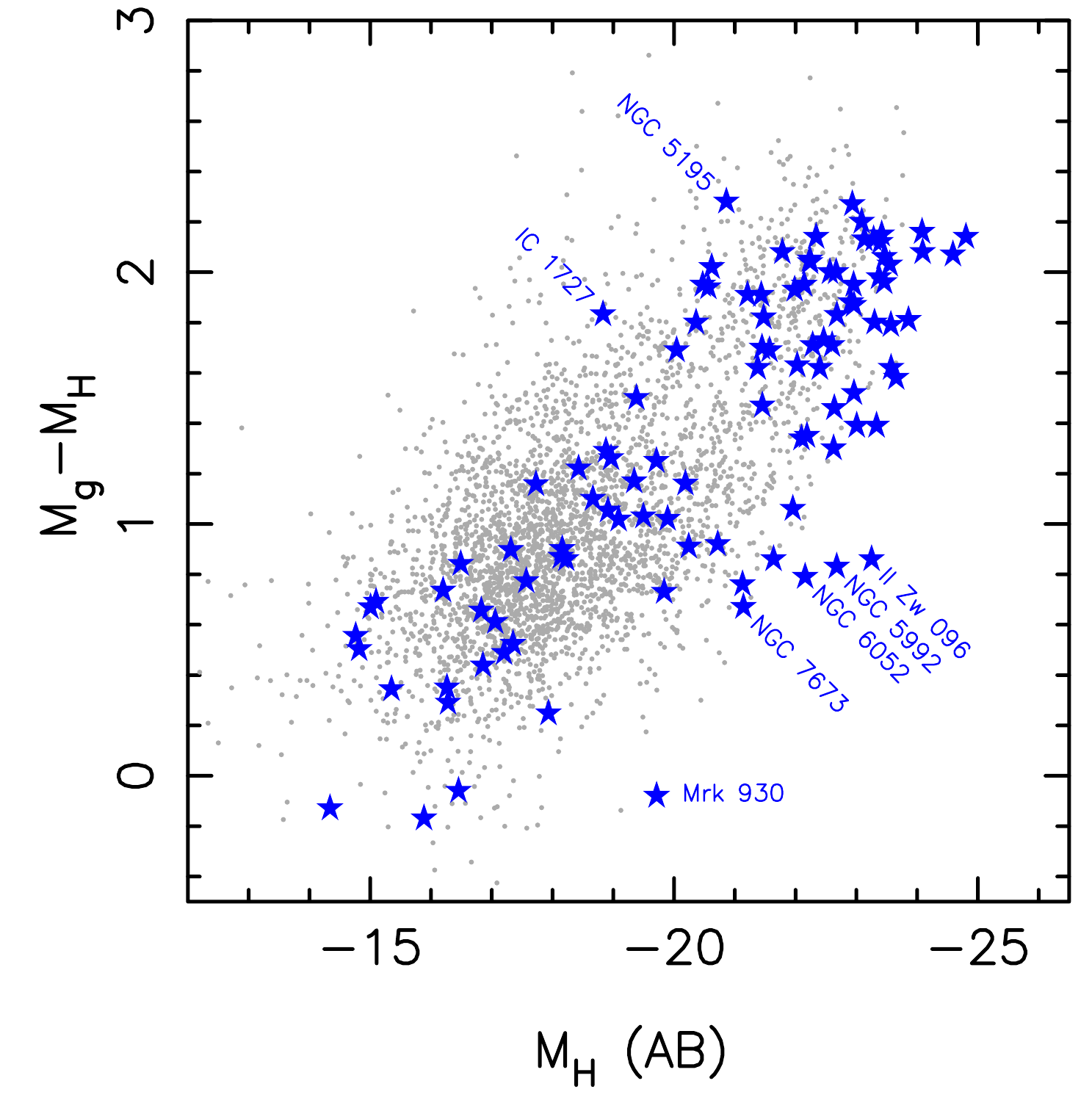}
\caption{The ultraviolet-infrared (left) and $g-H$ (right) color-magnitude relations for star-forming galaxies. Some of the outliers in the $g-H$ are labelled, and these are often merging galaxies rather than spirals. As the \citet{2014ApJS..212...18B,2017ApJ...847..136B} sample deliberately selected galaxies to span parameter space, it shows more scatter than the magnitude limited GAMA sample.  \label{fig:thefigure}}
\end{center}
\end{figure}

In Figure~\ref{fig:thefigure} we present the ultraviolet-infrared color-magnitude plot of $z\sim 0$ star-forming galaxies, using the $FUV$, $NUV$, and WISE $W3$ photometry. We find the best relation is produced when $A_{FUV}  = 2.6 (M_{FUV} - M_{NUV})$, which is shallower than the dust extinction relation of \citet{2011ApJ...741..124H}, where $A_{FUV}  = (3.83\pm 0.48) [M_{FUV} - M_{NUV} - (0.022\pm 0.024)]$. As a cross check of our results, in Figure~\ref{fig:thefigure} we also plot photometry of $z<0.05$ GAMA galaxies \citep{2016MNRAS.460..765W} with WISE $m_{W2}-m_{W3}>-0.5$, and we find good agreement although GAMA spans a smaller range of $M_{W3}$ than \citet{2014ApJS..212...18B, 2017ApJ...847..136B}. We also observe similar relations when we replace WISE $W3$ with WISE $W4$ or {\it Spitzer} $24~{\rm \mu m}$, albeit with more scatter. 

The best-fit color-magnitude relation is given by 
\begin{equation}
M_{W3} = -14.8 - 2.1 \times \left[ 2.6 (M_{FUV} - M_{NUV}) - M_{W3} \right].
\end{equation}
Using the  \citet{2014ApJS..212...18B, 2017ApJ...847..136B} sample, we find the $1\sigma$ scatter of $M_{W3}$ about the relation is $\sigma_{W3} =1.6~{\rm mag}$. If ultraviolet - infrared color was used as a distance indicator then the 68\% scatter of the distance would be a factor of $\sim 2$. 

We note color-magnitude relations for blue galaxies have been identified previously, including the median optical color of blue galaxies varying with magnitude \citep{2004ApJ...600..681B}. Furthermore, \citet{1982ApJ...257..527T} identified a tight $B-H$ color-magnitude relation for spiral galaxies, and in the right panel of Figure~\ref{fig:thefigure} we reproduce this relation for star-forming galaxies using SDSS $g$ and 2MASS $H$-band photometry.  \citet{1982ApJ...257..527T} attributed this relation to specific star formation rate, chemical abundances and/or initial mass function varying with mass. Interestingly, we do see some outliers in the $g-H$ versus $M_H$ diagram, including merging starbursts. These outliers are not unexpected, given $g$ and $H$ trace different galaxy properties, whereas the ultraviolet and mid-infrared are both (primarily) star formation rate tracers. 

In this note we have identified and characterized the ultraviolet-infrared color-magnitude relation of star-forming galaxies. The ultraviolet to mid-infrared flux ratios of star-forming galaxies span over two orders of magnitude and show a clear dependence on absolute magnitude from $M_{W3}\sim -13$ to $M_{W3}\sim -25$, which may present problems for models of galaxy spectral energy distributions that have been largely verified on $\sim L^*$ galaxies. The color-magnitude relation of star-forming galaxies illustrates the (broadband) spectral diversity of star-forming galaxies that results from established correlations between the physical properties and mass, including the mass-metallicity relation.
 

\end{document}